\def\pmb#1{\setbox0=\hbox{#1}%
\kern-.025em\copy0\kern-\wd0	
\kern-.05em\copy0\kern-\wd0
\kern-.025em\raise.0433em\box0}

\def \bomega {{\pmb{$\omega$}}}

\documentclass{elsart}
\usepackage{graphicx}

\usepackage{amssymb}

\begin{document}



\begin {frontmatter}
\title{ Magnetic Fields via Polarimetry: Progress of
Grain Alignment Theory}
\author[Alex]{A. Lazarian},
\address[Alex]{University of Wisconsin-Madison,
Astronomy Department, 475 N. Charter St., Madison, WI 53706, e-mail:
lazarian@astro.wisc.edu}



\begin{abstract}
Most astrophysical systems, e.g. stellar winds, the diffuse
interstellar 
medium, molecular clouds,
 are magnetized with magnetic fields that influence almost
all of 
their properties. One of the most informative techniques of magnetic
field studies is based on the use of starlight polarization and
polarized emission arising from aligned dust. How reliable
the interpretation of the polarization maps in terms of magnetic
fields is the issue that the grain alignment theory addresses. 
Although grain alignment is a problem of half a century standing,
recent progress achieved in the field makes us believe that we are 
approaching the solution of this mystery. I review basic physical
processes involved in grain alignment and discuss the niches for
different alignment mechanisms. I show why mechanisms that were favored for
decades do not look so promising right now, while the
radiative torque mechanism ignored for more than 20 years looks 
so attractive. I define the observational tests 
and outline the circumstances when grain alignment theory
predicts that new yet untapped information
of magnetic field structure is available through polarimetry.
In particular, I touch upon mapping magnetic fields in circumstellar
regions, interplanetary space and in comet comae. 
\end{abstract}
\end{frontmatter}

\section{Introduction}

Magnetic fields are of utmost importance most astrophysical systems.
Conducting matter is entrained on magnetic field lines and magnetic
pressure and tension are very important for its dynamics. 
For instance,
galactic magnetic fields play key role in many processes, including
star formation, mediating shocks, influencing heat and mass transport,
modifying turbulence etc. Aligned dust grains trace the magnetic field
and provide a unique source of information about magnetic field
structure. How reliable is this source of information? What are
the prospects of the polarimetric research? This review addresses
those questions while dealing with the problem of grain alignment
theory.

Grain alignment of interstellar dust 
has been discovered more than half a century ago. Hall (1949) and 
Hiltner (1949) reported polarization that 
was attributed to the differential extinction of starlight by
dust particles with longer axes preferentially aligned.
Very soon it was realized that the alignment happens with
respect to the interstellar magnetic field\footnote{The relation between
grain alignment direction and that of magnetic field is clear from
a comparison of synchrotron polarization maps and those of galactic starlight
polarization (see Serkowski, Mathewson \& Ford 1975). Recent measurements
of polarization in external galaxies (see Jones 2000) makes this relation
even more obvious.} Starting from that
moment polarized starlight and later the polarized emission
by aligned grains have become the principal technique of
studying magnetic field morphology in molecular clouds.
As magnetic fields are thought to control
star formation (see Savier, McKee \& Stahler 1997)
the value of the technique is difficult to overestimate.
However, to what extend the polarization maps trace magnetic fields
is a non-trivial question that the grain alignment theory deals with.

For many years grain alignment theory had a very limited predictive power
and was an issue of hot debates. 
This caused somewhat cynical approach to the theory among some of the
polarimetry practitioners who preferred to be guided in their work
by the following rules of thumb: {\it All grains are always aligned 
and the alignment happens with the longer grain axes perpendicular to
magnetic field.} This simple recipe was shattered, however, by observational
data which indicated that \\
I. Grains of sizes smaller than the critical size are either not aligned
or marginally aligned (Mathis 1986, Kim \& Martin 1995).\\
II. Carbonaceous grains are not aligned, but silicate grains are aligned
(see Mathis 1986).\\  
III. Substantial part of 
grains deep within molecular clouds are not aligned (Goodman et al. 1995,
Lazarian, Goodman \& Myers 1997).\\
VI. Grains might be aligned with longer axes parallel to 
magnetic fields\footnote{A simple, but not always clearly understood
property of grain alignment in interstellar medium is that it always 
happens  in respect to magnetic field. It can be shown that the
fast (compared with other time scales) Larmor
precession of grains makes the magnetic field the reference axis. 
Note, however.
that grains
 may align with their longer axes {\it perpendicular} or {\it parallel}
to magnetic field direction.
Similarly, magnetic fields may change their configuration and orientation
in space (e.g. due to Alfven waves), but if the time for such
a change is much longer than the Larmor period the alignment of
grains {\it in respect to the field lines} persists as the consequence
of preservation of the adiabatic invariant.} (Rao et al 1998).

These facts could persuade even the most stubborn types 
that the interpretation
of interstellar polarimetric data does require an adequate 
theory. A further boost of the interest to 
grain alignment  came from the
search of Cosmic Microwave Background (CMB) polarization (see
Lazarian \& Prunet 2002, for a review). Aligned dust in this case acts as
a source of a ubiquitous 
foreground that is necessary to remove from the data. It is clear
that understanding of grain alignment is the key element for such a removal.

With the present level of interest to the CMB polarization 
we are bound to have a lot of microwave and far infrared polarimetry
data. It is important to understand to what extend this data reflects
the structure of magnetic field in the Galaxy and whether this
data can be used to get insight into the processes of galactic magnetic
field generation and into interstellar turbulence\footnote{Velocity
and magnetic field statistics provide the most clear insight in
what is going on with the turbulence. With velocity statistics available
through the recently developed Velocity Channel Analysis (VCA) technique (
Lazarian \& Pogosyan 2000) magnetic fields statistics is the missing
element. Polarized starlight and emission from aligned grains
provide the easiest way to get such a statistics.}. 

While the alignment of interstellar dust is a generally accepted fact, 
the alignment of
dust in conditions other than interstellar has not been fully appreciated.
The common explanation of light polarization from comets or circumstellar
regions is based on light scattering by randomly oriented particles. 
The low efficiency and slow rates
of alignment were quoted to justify such an approach (see Bastien
1988). This point of view is common in spite of the mounting
evidence in favor of grain alignment (see Briggs \& Aitken 1986,
Aitken et al. 1995, Tamura et al. 1995). 
However, recent advances in understanding of grain alignment show
that it is an efficient and rapid process. Therefore, we
do expect to have circumstellar, interplanetary and comet 
dust aligned. This
opens new exciting avenues for polarimetry.  

Traditionally linear starlight
polarimetry was used. These days 
far infrared polarimetry of dust
emission has become
the major source of molecular field structure data (see Hildebrand 2000).
It is possible that circular polarization may become an important
means of probing magnetic fields in circumstellar
regions and comets.

In this review I claim that the modern grain alignment 
theory allows us to solve most of the existing puzzles and 
can be used successfully to interpret polarimetry in terms of magnetic
field.
A substantial part of the review is devoted to the 
physics of grain alignment, which 
 is deep and exciting. It is enough to say that
its study resulted in a discovery of a few new solid state effects. 
The rich physics of grain alignment (see Fig~1a for an illustration
of motion complexity) presents a problem, however,
for its presentation. Therefore I shall describe first the genesis
of ideas that form the basis of the present-day grain alignment theory.
The references to the original papers should help the interested reader
to get the in-depth coverage of the topic. Earlier reviews on the
subject include Hildebrand (1988), Roberge (1996), Lazarian, Goodman
\& Myers (1997), Lazarian (2000). 

In what follows we show how the properties of polarized radiation
is related to the statistics of aligned grains (section~2), analyze
the major alignment mechanisms (section~3), discuss observational data
that allows to distinguish between different alignment processes (section~4)
and outline the prospects of using grain alignment to study circumstellar,
interplanetary magnetic fields (section~5). A discussion and summary are
provided in sections~6 and 7.

\section{Aligned Grains \& Polarized Radiation}

\subsection{Linear Polarized Starlight from Aligned Grains}

For an ensemble of aligned grains the extinction perpendicular 
and parallel to
the direction
of alignment and parallel are different\footnote{According
to Hildebrand \& Dragovan (1995) the best fit of the grain properties
corresponds to oblate grains with the ratio of axis about 2/3.}. Therefore
 that is initially unpolarized starlight acquires polarization while
passing through a volume with aligned grains.
If the extinction in the direction of alignment is $\tau_{\|}$ and in
the perpendicular direction is  $\tau_{\bot}$
one can write the polarization, $P_{abs}$, by selective extinction
 of grains
as 
\begin{equation}
P_{abs}=\frac{e^{-\tau_{\|}}-e^{-\tau_{\bot}}}{e^{-\tau_{\|}}+e^{-\tau_{\bot}}}
\approx -{(\tau_{\|}-\tau_{\bot})}/2~,
\label{Pabs}
\end{equation}
where the latter approximation is valid for $\tau_{\|}-\tau_{\bot}\ll 1$.
To relate the difference of extinctions to the properties of aligned grains
one can take into 
account the fact that the extinction is proportional to the product
of the grain density and  their cross sections. If a cloud is composed of 
identical aligned grains
$\tau_{\|}$ and $\tau_{\bot}$ are proportional to the number of grains
along the light path times the corresponding cross sections, which
are, respectively, 
$C_{\|}$ and $C_{\bot}$.

In reality one has to consider additional complications like
incomplete grain alignment, and variations in the direction
of the alignment axis (in most cases the latter is the direction of
 magnetic field, as discussed above) along the line of sight.  
 To obtain 
an adequate description one can (see Roberge \& Lazarian 1999) consider
 an electromagnetic wave propagating along the line of sight
{\mbox{$\hat{\bf z}^{\bf\rm o}$}} axis.
The transfer equations for the Stokes parameters
depend on the cross sections,  $C_{xo}$ and $C_{yo}$, for linearly polarized
waves with the electric vector,  {\mbox{\boldmath$E$}}, 
along the {\mbox{$\hat{\bf x}^{\bf\rm o}$}} and 
{\mbox{$\hat{\bf y}^{\bf\rm o}$}} directions
that are in the plane perpendicular to {\mbox{$\hat{\bf z}^{\bf\rm o}$}}
(see Lee \& Draine 1985).

To calculate  $C_{xo}$ and $C_{yo}$,
one transforms the components of {\mbox{\boldmath$E$}} to
a frame aligned with the principal axes of the grain and
takes the appropriately-weighted sum of the
cross sections, $C_{\|}$ and , $C_{\bot}$ for {\mbox{\boldmath$E$}}
 polarized along the grain
axes (Fig~1b illustrates the geometry of observations).
When the transformation is carried out and the resulting
expressions are averaged over precession angles, one finds (see
transformations in Lee \& Draine 1985 for spheroidal grains and in
Efroimsky 2002 for a general case)
that
the mean cross sections are
\begin{equation}
C_{xo} = C_{avg} + \frac{1}{3}\,R\,\left(C_{\bot}-C_{\|}\right)\,
       \left(1-3\cos^2\zeta\right)~~~,
\label{eq-2_5}
\end{equation}
\begin{equation}
C_{yo} = C_{avg} + \frac{1}{3}\,R\,\left(C_{\bot}-C_{\|}\right)~~~,
\label{eq-2_6}
\end{equation}
where $\zeta$ is the angle between the polarization axis and the 
{\mbox{$\hat{\bf x}^{\bf\rm o}$}} {\mbox{$\hat{\bf y}^{\bf\rm o}$}}
plane;
$C_{avg}\equiv\left(2 C_{\bot}+ C_{\|}\right)/3$ is the effective
cross section for randomly-oriented grains.
To characterize the alignment we used in eq.~(\ref{eq-2_6})
 the Raylegh reduction factor
(Greenberg 1968)
\begin{equation}
R\equiv \langle G(\cos^2\theta) G(\cos^2\beta)\rangle
\label{R}
\end{equation}
where angular brackets denote ensemble averaging, $G(x) \equiv 3/2 (x-1/3)$,
 $\theta$ is the angle between the axis of the largest moment of inertia
(henceforth the axis of maximal inertia) and the magnetic field ${\bf B}$, while
$\beta$ is the angle between the angular momentum ${\bf J}$ and ${\bf B}$. 
 To characterize
${\bf J}$ alignment in grain axes and in respect to magnetic field,
 the
measures ${Q_X\equiv \langle G(\theta)\rangle}$ and 
$Q_J\equiv \langle G(\beta)\rangle$
are used.
Unfortunately, these statistics 
are not independent and therefore $R$ is not equal to $Q_J Q_X$ (see 
Lazarian 1998, Roberge
\& Lazarian 1999). This considerably complicates 
the treatment of grain alignment.

\subsection{Polarized Emission from Aligned Grains}

The difference in $\tau_{\|}$ and $\tau_{\bot}$ results in emission of
aligned grains being polarized:
\begin{equation}
P_{em}=\frac{(1-e^{-\tau_{\|}})-(1-e^{-\tau_{\bot}})}{(1-e^{-\tau_{\|}})+
(1-e^{-\tau_{\bot}})}\approx \frac{\tau_{\|}-\tau_{\bot}}
{\tau_{\|}+\tau_{\bot}}~,
\label{Pem}
\end{equation}
where both the optical depths $\tau{\|}$ are $\tau_{\bot}$ were
assumed to be small. Taking into account that 
both $P_{em}$ and $P_{abs}$ are functions
of wavelength $\lambda$ and 
combining eqs.(\ref{Pabs}) and (\ref{Pem}), one gets for
$\tau=(\tau_{\|}+\tau_{\bot})/2$
\begin{equation}
P_{em}(\lambda) \approx -P_{abs}(\lambda)/\tau(\lambda)~,
\label{Pem}
\end{equation}
which establishes the relation between polarization in emission and
absorption. The minus sign
in eq~(\ref{Pem})
reflects the fact that emission and absorption polarization are
orthogonal. 
As $P_{abs}$ depends on $R$, $P_{em}$ also depends on the Rayleigh reduction
factor.

\subsection{Circular Polarization from Aligned Grains}

A way of obtaining circular polarization is to have a magnetic field
that varies along the line of sight (Martin 1972). Passing through one
cloud with aligned dust the light becomes partially linearly polarized.
On passing the second cloud with dust aligned in a different direction the
light gets circular polarization.
Literature study shows that this  effect that is well remembered 
(see Menard et al 1988), while the other process that also creates
circular polarization is frequently forgotten. We mean the process
of single scattering of light on aligned particles. Electromagnetic
wave interacting with a single grain coherently excites dipoles parallel
and perpendicular to the grain long axis. In the presence of adsorption
these dipoles get phase shift giving rise to circular polarization. 
This polarization can be observed from the ensemble of grains if
the grains are aligned. The intensity of circularly polarized
 component of radiation emerging via scattering of radiation with
${\bf k}$ wavenumber on small ($a\ll \lambda$) spheroidal
particles is (Schmidt 1972)
\begin{equation}
V( {\bf e}, {\bf e}_0, {\bf e}_1)=\frac{I_0 k^4}{2 r^2}i(\alpha_{\|}
\alpha^{\ast}_{\bot}-\alpha^{\ast}_{\|}\alpha_{\bot})\left([{\bf e_0}\times
{\bf e}_1] {\bf e}\right)({\bf e}_0 {\bf e}),
\end{equation}
where ${\bf e}_0$ and ${\bf e}_1$ are the unit vectors in the directions
of incident and scattered radiation, ${\bf e}$ is the direction along
aligned axes of spheroids; $\alpha_{\bot}$ and $\alpha_{\|}$ are particle
polarizabilities along ${\bf e}$ and perpendicular to it.

The intensity of the circulary polarized radiation scattered in the
volume $\Delta \Gamma({\bf d}, {\bf r})$ at $|{\bf d}|$ from the star
and distances $|{\bf r}|$ from the observer is (Dolginov \& Mytrophanov 1978)
\begin{equation}
\Delta V ({\bf d}, {\bf r})=\frac{L_{\star} n_{\rm dust}\sigma_{V}}{6\pi |{\bf d}|^4
|{\bf r}||{\bf d}-{\bf r}|^2}R \left([{\bf d}\times {\bf r}] h\right)
({\bf d}{\bf r})\Delta \Gamma({\bf d}, {\bf r})~~~,
\label{circular}
\end{equation}
where $L_{\star}$ is the stellar luminosity, $n_{\rm dust}$ is number 
density of dust grains and $\sigma_V$ is the cross section for
producing circular polarization, which is for small grains
is $\sigma_V=i/(2k^4)(\alpha_{\|}\alpha^{\ast}_{\bot}-\alpha^{\ast}_{\|}\alpha_{\bot})$.
According to Dolginov \& Mytrophanov (1978) circular polarization arising 
from single scattering on aligned grains
can be as high as several percent for metallic or graphite particles,
which is much more than one may expect
from varying magnetic field direction along the line of
sight (Martin 1972). In the latter case linear polarization produced
by one layer of aligned grains passes through another layer where
alignment direction is different. If passing through a single
layer the linear polarization degree is $p$, passing through two
layers produces circular polarization that does not exceed $p^2$. 

\section{Grain Alignment Theory: New and Old Ideas}

We have seen in the previous sections that both linear and circular
polarizations depend on the degree of grain alignment given by
$R$-factor (\ref{R}). Therefore it is the goal of grain alignment theory
to determine this factor. The complexity of the grain alignment is
illustrated in Fig~1, which shows that grain alignment is indeed
a multi-stage process.

A number of different mechanisms that produce grain alignment has been
developed by now (see table~1 in Lazarian, Goodman \& Myers 1997). Dealing
with a particular situation one has to identify the dominant alignment process.
Therefore it is essential to understand different mechanisms. By now
the theory of grain alignment is rather complex. This makes it advantageous
to follow the evolution of grain alignment ideas. It is instructive
to see the major role that observations played in shaping up of the theory.
    
\subsection{Foundations of the Theory}

The first stage of alignment theory development
started directly after the discovery of
starlight polarization. 
Nearly simultaneously Davis \& Greenstein (1950) 
and Gold (1951) proposed their scenarios of alignment. 

{\it Paramagnetic Alignment: Davis-Greenstein Process}\\
Davis-Greenstein
mechanism (henceforth D-G mechanism)
is based on the paramagnetic dissipation that is experienced
by a rotating grain. Paramagnetic materials contain unpaired
electrons which get oriented by the interstellar magnetic field ${\bf B}$. 
The orientation of spins causes
grain magnetization and the latter 
varies as the vector of magnetization rotates
 in grain body coordinates. This causes paramagnetic loses 
at the expense of grain rotation energy.
Note, that if the grain rotational velocity ${\bomega}$
is parallel to ${\bf B}$, the grain magnetization does not change with time
and therefore
no dissipation takes place. Thus the
paramagnetic dissipation  acts to decrease the component of ${\bomega}$
perpendicular to ${\bf B}$ and one may expect that eventually
grains will tend to rotate with ${\bomega}\| {\bf B}$
provided that the time of relaxation $t_{D-G}$ is much shorter than  $t_{gas}$,
the
time of randomization through chaotic gaseous bombardment.
In practice, the last condition is difficult to satisfy. For $10^{-5}$ cm 
grains
in the diffuse interstellar medium
$t_{D-G}$ is of the order of $7\times 10^{13}a_{(-5)}^2 B^{-2}_{(5)}$s , 
while  $t_{gas}$ is $3\times 10^{12}n_{(20)}T^{-1/2}_{(2)} a_{(-5)}$ s (
see table~2 in Lazarian \& Draine 1997) if
magnetic field is $5\times 10^{-6}$ G and
temperature and density of gas are $100$ K and $20$ cm$^{-3}$, respectively. 
However, in view of uncertainties in
interstellar parameters the D-G theory initially looked plausible.

{\it Mechanical Alignment: Gold Process}\\
Gold mechanism is a process of mechanical alignment of grains. Consider
a needle-like grain interacting with a stream of atoms. Assuming
that collisions are inelastic, it is easy to see that every
bombarding atom deposits angular momentum $\delta {\bf J}=
m_{atom} {\bf r}\times {\bf v}_{atom}$ with the grain, 
which is directed perpendicular to both the
needle axis ${\bf r}$ and the 
 velocity of atoms ${\bf v}_{atom}$. It is obvious
that the resulting
grain angular momenta will be in the plane perpendicular to the direction of
the stream. It is also easy to see that this type of alignment will
be efficient only if the flow is supersonic\footnote{Otherwise grains
will see atoms coming not from one direction, but from a wide cone of
directions (see Lazarian 1997a) and the efficiency of 
alignment will decrease.}.
Thus the main issue with the Gold mechanism is to provide supersonic
drift of gas and grains. Gold originally proposed collisions between
clouds as the means of enabling this drift, but later papers (Davis 1955) 
showed that the process could  only align grains over limited patches of
interstellar space, and thus the process
cannot account for the ubiquitous grain 
alignment in diffuse medium.

{\it Quantitative Treatment and Enhanced Magnetism}\\
The first detailed analytical treatment of the problem of D-G
alignment was given by Jones \& Spitzer (1967) who described the alignment
of ${\bf J}$
using a Fokker-Planck equation. This 
approach allowed them to account for magnetization fluctuations
within grain material and thus provided a more accurate picture of 
${\bf J}$ alignment.
$Q_X$ was assumed to follow
the Maxwellian distribution, although the authors noted
that this might not be correct. 
The first numerical treatment of
D-G alignment was presented by Purcell (1969). 
By that time it became clear that the D-G
mechanism is too weak to explain the observed grain alignment. However,
Jones \& Spitzer (1969) noticed that if interstellar grains
contain superparamagnetic, ferro- or ferrimagnetic (henceforth SFM) 
inclusions\footnote{The evidence for such inclusions was found much later
through the study of interstellar dust particles captured in
the atmosphere (Bradley 1994).}, the
$t_{D-G}$ may be reduced by orders of magnitude. Since $10\%$ of
atoms in interstellar dust are iron
the formation of magnetic clusters in grains was not far fetched
(see Spitzer \& Turkey 1950, Martin 1995)
and therefore the idea was widely accepted. Indeed, with enhanced 
magnetic susceptibility the D-G mechanism was able to solve
all the contemporary problems of alignment. The conclusive
at this stage was the paper by Purcell \& Spitzer (1971) where
all various models of grain alignment, including, for
instance, the model of cosmic ray alignment 
by Salpeter \& Wickramasinche (1969)
and photon alignment by Harwit (1970)  
were quantitatively discussed and the D-G model with enhanced
magnetism was endorsed. It is this stage of development that is widely
reflected in many textbooks.

\subsection{Additional Essential Physics}

{\it Barnett Effect and Fast Larmor Precession}\\
It was realized by Martin (1971) that rotating charged grains will develop
magnetic moment and the interaction of this moment with the interstellar
magnetic field will result in grain precession. The characteristic
time for the precession was found to be comparable with $t_{gas}$. 
However, soon  a process that
renders much larger magnetic moment was discovered (Dolginov \& Mytrophanov 
1976). This process is the 
Barnett effect, which is converse of the Einstein-de Haas effect.
If in Einstein-de Haas effect a paramagnetic body starts rotating
 during remagnetizations
as its flipping 
electrons transfer the angular momentum (associated with their spins)
 to the
lattice, in the Barnett effect
the rotating body shares its angular momentum with the electron
subsystem  causing magnetization. The magnetization
is directed along the grain angular velocity and the value
of the Barnett-induced magnetic moment is $\mu\approx 10^{-19}\omega_{(5)}$~erg
gauss$^{-1}$ (where $\omega_{(5)}\equiv \omega/10^5{\rm s}^{-1}$). Therefore
the Larmor precession has a period $t_{Lar}\approx 3\times 10^6 B_{(5)}^{-1}$~s and 
the magnetic field defines the axis of alignment as we explained in section~1.

{\it Suprathermal Paramagnetic Alignment: Purcell Mechanism}\\
The next step was done by Purcell(1975, 1979),
who discovered that grains can rotate much faster than were previously
thought. He noted 
that variations of photoelectric yield, the H$_2$ formation efficiency,
and  variations of accommodation coefficient over grain surface
would result in uncompensated torques acting upon a
grain. The H$_2$ formation on the grain surface clearly illustrates the
process we talk about: if H$_2$ formation takes place only over particular
catalytic sites, these sites act as miniature rocket engines
spinning up the grain. Under such uncompensated torques the grain will spin-up to
velocities much higher than thermal (Brownian) and Purcell termed those
velocities ``suprathermal''. Purcell also noticed that for suprathermally
rotating grains
internal relaxation will bring ${\bf J}$ parallel to the axis of maximal
inertia (i.e. $Q_X=1$). Indeed, for an oblate spheroidal
grain with 
angular momentum ${\bf J}$  the energy can be written
\begin{equation}
E(\theta)=\frac{J^2}{I_{max}}\left(1+\sin^2\theta (h-1)\right)
\label{e}
\end{equation}
where $h=I_{max}/I_{\bot}$ is the ratio of the maximal to minimal moments
of inertia. Internal forces cannot change the angular momentum, but
it is evident from Eq.~(\ref{e}) that the energy can be decreased by
aligning the axis of maximal inertia along ${\bf J}$, i.e. decreasing
$\theta$. Purcell (1979) discusses two possible causes of internal 
dissipation,
the first one related to the well known inelastic relaxation, the second is
due to the mechanism that he discovered and termed ``Barnett relaxation''.
This process may be easily understood. We know that a 
freely rotating grain preserves the direction of
${\bf J}$, while angular velocity precesses about 
${\bf J}$ and in grain body axes.
We learned earlier that the Barnett effect results in the magnetization
vector parallel to $\bomega$. As a result, the Barnett magnetization
will precess in body axes and cause paramagnetic relaxation.
The ``Barnett equivalent magnetic field'', i.e. the equivalent external
magnetic field that would cause the same magnetization of the grain  
material, is $H_{BE}=5.6 \times10^{-3} \omega_{(5)}$~G, 
which is much larger than the interstellar magnetic 
field. Therefore the Barnett relaxation happens on the scale $t_{Bar}\approx
4\times 10^7 \omega_{(5)}^{-2}$~sec,
i.e. essentially instantly compared to $t_{gas}$ and $t_{D-G}$.

{\it Theory of Crossovers}\\
If $Q_X=1$ and the suprathermally rotating grains are immune to randomization
by gaseous bombardment, will paramagnetic grains be perfectly aligned with
$R=1$? This question was addressed by Spitzer \& McGlynn (1979)
(henceforth
SM79) who observed
that adsorption of heavy elements on a grain  should result
in the resurfacing phenomenon that, e.g.  should remove early sites
of H$_2$ formation and create new ones. As the result,
H$_2$ torques will occasionally change their direction and spin the grain
down. SM79  showed that in the absence of
random torques the spinning down grain will
 flip over preserving the direction of its original angular momentum.
However, in the presence of random torques
 this direction will be altered with the maximal deviation inflicted
over a short period of time just before and after the flip, i.e.
during the time when the value of grain angular momentum is minimal.
The actual value of angular momentum during this critical
period depends on the ability of ${\bf J}$ to deviate from
the axis of maximal inertia.
SM79 observed that as the Barnett relaxation 
couples ${\bf J}$ with the axis of maximal inertia it 
makes randomization of grains during crossover nearly complete. With the
resurfacing time $t_{res}$ estimated by SM79 to be of the order of $t_{gas}$,
the gain of the alignment efficiency was
insufficient to reconcile the theory and observations unless the grains 
had SFM inclusions.

{\it Radiative Torques}\\
If the introduction of the concept of suprathermality by Purcell changed
the way researchers thought of grain dynamics, the introduction of radiative torques
passed essentially unnoticed. Dolginov (1972) argued that quartz grains
may be spun up due to their specific rotation of polarization
while later Dolginov \& Mytrophanov (1976)
discovered that irregular grain shape may allow grains scatter left and right
hand polarized light differentially,  thus spinning up helical grains through
scattering of photons\footnote{The principal difference between radiative
torque mechanism and the radiative emission/absorption mechanism proposed
by Harwit (1970) is that the radiative torques are regular and therefore
increase the grain velocity in proportion to time. Harwit's mechanism,
on the other hand, is based on stochastic spin-up and therefore is
subdominant. We also note that the emission of photons is insensitive
to grain helicity as the emitted photons have wavelengths much larger than
the grain size.}. They stressed that the most efficient spin-up
is expected when grains size is comparable with the wavelength and estimated
the torque efficiency for particular helical grain shapes, but failed
to provide estimates of the relative efficiency of the mechanism in the
standard interstellar conditions. In any case, this ingenious idea had not
been appreciated for another 20 years.

{\it Observational tests: Serkowski Law}\\
All in all, by the end of seventies the the following alignment mechanisms
were known:
1. paramagnetic( a. with SFM inclusions,
   b. with suprathermal rotation),
2. mechanical,
3. radiative torques.
The third was ignored, the second was believed to be suppressed
for suprathermally rotating grains, which left
the two modifications of the paramagnetic mechanism as competing alternatives.
Mathis (1986) noticed that the interstellar polarization-wavelength dependence
known as the Serkowski law (Serkowski et al. 1975) can be explained if
grains larger that $\sim 10^{-5}$~cm are aligned, while smaller grains
are not. To account for this behavior Mathis (1986) noticed
that the SFM inclusions will have a better chance to
be in larger rather than smaller grains. The success of fitting
observational data persuaded
the researchers that the problem of grain alignment is solved at last.

\subsection{Present Stage of Grain Alignment Theory}

Optical and near infrared observations by 
Goodman et al. (1995)  showed 
that polarization efficiency may
drop within dark clouds while far infrared observations by Hildebrand et al. (1984),
Hildebrand et al. (1990)
revealing aligned grains within star-forming dark clouds. This
renewed interest to grain alignment problem.

{\it New Life of Radiative Torques}\\
Probably the most dramatic change of the picture was the unexpected advent
of radiative torques. Before Bruce Draine realized that the torques
can be treated with the versatile discrete dipole approximation (DDA)
code ( Draine \& Flatau 1994), their role was unclear. For instance, earlier on
difficulties associated with the analytical approach to
the problem were discussed in Lazarian (1995a).
However, very soon after that Draine (1996) modified the DDA code
 to calculate the torques acting on grains of arbitrary
shape. His work revolutionized the field! 
The magnitude of torques were found to be substantial and present
for grains of various irregular shape (Draine 1996, Draine \& Weingartner
1996). After that it became impossible
to ignore these torques. Being related to grain shape, rather than surface
these torques are long-lived\footnote{In the case of the Purcell's
rockets the duration of torque action is limited by the time of resurfacing,
while in the case of radiative torques it is the time scale on which the
grain is either destroyed via collisions, coagulates with another grain
or gets a different shape in the process of growth.}, i.e. 
$t_{spin-up}\gg t_{gas}$, 
which allowed Draine \& Weingartner (1996)
to conclude that in the presence of isotropic radiation the radiative 
torques can support fast grain rotation long enough in order for
paramagnetic torques to align grains (and without any SFM
inclusions). However, the important question was what would happen
in the presence of anisotropic radiation. Indeed, in the presence
of such radiation the torques will change as the grain aligns
 and this may result in a spin-down. Moreover,
anisotropic flux of radiation will deposit angular momentum 
which is likely to overwhelm rather weak paramagnetic torques. These sort of
questions were addressed by Draine \& Weingartner (1997) and it was
found that for most of the tried grain shapes the torques tend to 
align ${\bf J}$ along magnetic field. The reason for that is yet unclear
and some caution is needed as the existing treatment ignores the dynamics
of crossovers which is  very important for the alignment of
suprathermally rotating grains. Nevertheless, radiative torques
are extremely appealing as their predictions are consistent
with observational data (see Lazarian, Goodman \& Myers 1997, Hildebrand et 
al. 1999, see section 4 as well).

{\it New Elements of Crossovers}\\
Another unexpected development was a substantial change of the picture
of crossovers. As we pointed out earlier,  Purcell's discovery of
fast internal dissipation resulted in a notion that ${\bf J}$
should always stay along the axis of maximal inertia as long
as $t_{dis}\ll t_{gas}$. Calculations in
SM79 were based on this notion.
However, this
perfect coupling
 was questioned in Lazarian (1994) (henceforth L94), where it was shown that
thermal fluctuations within grain material partially randomize
the distribution of grain axes in respect to ${\bf J}$. 
The process was quantified in Lazarian \& Roberge (1997)
(henceforth LR97),
where the distribution of $\theta$ for a freely
rotating grain was defined through the Boltzmann distribution
 $\exp(-E(\theta)/kT_{grain})$,
where the energy $E(\theta)$ is given by Eq.~(\ref{e}). This finding
changed the understanding of crossovers a lot. First of all,
Lazarian \& Draine (1997)(henceforth LD97) observed that  thermal
fluctuations partially decouple ${\bf J}$ and the axis of maximal
inertia and therefore the value of angular moment at the moment
of a flip is substantially larger than SM79 assumed. Thus the
randomization during a crossover is  reduced and  LD97 obtained
a nearly
perfect alignment  for interstellar grains
rotating suprathermally, provided that
the grains were larger than a certain critical size $a_c$.  The latter 
size was found by
equating the time of the crossover and the time of the internal
dissipation $t_{dis}$. For $a<a_c$
Lazarian \& Draine (1999a) found new physical effects, which they termed
``thermal flipping'' and ``thermal trapping''. The thermal flipping
 takes place
as the time of the crossover becomes larger than $t_{dis}$.           
In this situation thermal fluctuations will enable flipovers. However,
being random, thermal fluctuations are likely to produce not a single
flipover, but multiple ones. As the grain flips back and forth the
regular (e.g. H$_2$) torques average out and the
grain can spend a lot of time rotating with thermal velocity, i.e.
being ``thermally trapped''. The paramagnetic alignment of 
grains rotating with 
thermal velocities is small (see above) 
and therefore grains with $a<a_{c}$ are
expected to be marginally aligned. The picture of preferential
alignment of large grains, as we know, corresponds to the Serkowski
law and therefore the real issue is to find the value of $a_c$.
The Barnett relaxation\footnote{A study by Lazarian \& Efroimsky (1999)
corrected the earlier estimate by Purcell (1979), but left the conclusion
about the Barnett relaxation
dominance, and therefore the value of $a_c$, intact. For larger objects, e.g.
for astreroids, comets, the inelastic relation is dominant (Efroimsky 
\& Lazarian 2000, Efroimsky 2001).}
 provides a comforting value of $a_c\sim 10^{-5}$~cm. However, in a
recent paper Lazarian \& Draine (1999b) reported a new solid state effect
that they termed ``nuclear relaxation''. This is an analog of Barnett
relaxation effect that deals with nuclei. Similarly to unpaired electrons
nuclei tend to get oriented in a rotating body. However the nuclear analog
of ``Barnett equivalent'' magnetic field is much larger and Lazarian \&
Draine (1999) concluded that the nuclear relaxation can be a million times
faster than the Barnett relaxation. If this is true $a_c$ becomes of the
order $10^{-4}$~cm, which means that the majority of interstellar grains
undergo constant flipping and rotate essentially thermally in spite of
the presence of
uncompensated Purcell torques. The radiative torques
are not fixed in body coordinates and it is likely that they can provide
a means for suprathermal rotation for grains that are larger than the
wavelength of the incoming radiation. Naturally, it is of utmost importance
to incorporate the theory of crossovers into the existing codes, 
and this work is under way.

{\it New Ideas and Quantitative Theories}\\
An interest to grain alignment resulted in search of new mechanisms. For
instance, Sorrell (1995a,b) proposed a mechanism of grain spin-up due to
interaction with cosmic rays that locally heat grains and provide evaporation
of adsorbed H$_2$ molecules. However, detailed
calculations in Lazarian \& Roberge (1997b)
showed that the efficiency of the torques was overestimated; the observations
(Chrysostomou et al. 1996) did not confirm Sorrell's predictions either. 
A more promising idea that
ambipolar diffusion can align interstellar grains was
put forward in Roberge \& Hanany (1990)(calculations are done
in Roberge et al. 1995). Within this mechanism ambipolar
drift provides the supersonic velocities necessary for mechanical alignment.   
Independently L94 proposed a mechanism of mechanical grain alignment
using Alfven waves. Unlike the ambipolar diffusion, this mechanism 
operates even in ideal MHD and relies only on the difference in inertia of
atoms and grains and on the direct interaction of grains with fluctuating
magnetic field (Lazarian \& Yan 2002, Yan \& Lazarian 2002). 
An additional boost to interest to mechanical
processes was gained when it was shown that suprathermally rotating
grains can be aligned mechanically (Lazarian 1995, Lazarian \& Efroimsky 1996,
Lazarian, Efroimsky \& Ozik 1996, Efroimsky 2002).
As it was realized that thermally rotating grains do not ${\bf J}$ tightly
coupled with the axis of maximal inertia (L94) and the effect
was quantified (LR97), it got possible to formulate quantitative theories
of Gold (Lazarian 1997a) and Davis-Greenstein (Lazarian 1997b, Roberge \& 
Lazarian 1999) alignments. Together with a better understanding of
grain superparamagnetism (Draine \& Lazarian 1998a),
damping of grain rotation (Draine \& Lazarian 1998b) and resurfacing of
grains (Lazarian 1995c), these developments
increased the predictive power of the grain alignment theory.

{\it Alignment of PAH}\\
All the studies above dealt with classical ``large'' grains. What  about
very small (e.g. $a<10^{-7}$~cm) grains? Can they be aligned? The answer
to this question became acute after Draine \& Lazarian (1998) explained
the anomalous galactic emission in the range $10-100$~GHz as arising
from rapidly (but thermally!) 
spinning tiny grains. This rotational dipole emission will
be polarized if grains are aligned. Lazarian \& Draine (2000) (henceforth LD00)
 found
that the generally accepted picture of the D-G relaxation is incorrect
when applied to such rapidly rotating ($\omega > 10^8$~s$^{-1}$) particles. 
Indeed, the D-G mechanism assumes
that the relaxation rate is the same whether grain
rotates in stationary magnetic field or magnetic field rotates around
a stationary grain. However, as grain rotates, we know that it gets
magnetized via Barnett effect  and it is known that
the relaxation rate within a magnetized body differs
from that in unmagnetized body. A non-trivial finding in LD00
was that the Barnett magnetization provides the 
optimal conditions for the paramagnetic
relaxation which enables grain alignment at frequencies for which the D-G
process is quenched (see Draine 1996). 
LD00 termed the process ``resonance relaxation'' to
distinguish from the D-G process and calculated the expected alignment values
for grains of different sizes. Will this alignment be seen through
infrared emission of small transiently heated small grains (e.g. PAH)?
The answer is probably negative\footnote{Earlier calculations by
Rouan et al (1992) of the problem were done at a time when
much of the relevant physics, e.g. resonance relaxation, thermal flipping,
incomplete internal relaxation were not known.}. The reason for
the such an answer is that internal
alignment of ${\bf J}$ and the axis of maximal inertia is being essentially
destroyed if a grain is heated up to high temperatures (LR97).
Therefore even if ${\bf J}$ vectors are well
aligned, grain axes will wobble with large amplitude about ${\bf B}$.
The expected alignment in terms of $R$ and therefore the
 polarization of emitted infrared photons, will be marginal in agreement
with observations (Sellgren, Rouan \& Leger 1988).

\section{Observational Tests of Interstellar Alignment}

As the reader may see from the previous discussion that
that several times the problem of grain alignment
seemed to be solved. However, the accumulation
of new observational facts and deeper insights into grain physics
caused  the changes of paradigms. The problem was attacked again
and again at a higher level of understanding. Three important
questions arise:\\
I. Do we need to keep in mind different mechanisms while dealing
with data?\\
II. Which mechanism dominates in which environment?\\
III. Are we still missing essential physics?

The answer to the first question is positive. As pointed out by
Hildebrand (1988) astrophysical environments present such a variety
of conditions that it is likely that every mechanism has
its own niche. How wide is this niche depends on the special conditions
required for the mechanism operation as well as its efficiency.
This brings us to the second and third questions. As we shall see below,
the present day theory provides quantitative predictions that
can be tested. So far this tests are consistent with the theoretical
predictions. More studies, both observational and theoretical,
are necessary, however.

\subsection{Testing Alignment in Molecular Clouds}
 
The data on grain alignment in molecular clouds looked at some point
very confusing. On one hand, optical and near-infrared polarimetry
of background stars did not show an increase of polarization degree 
with the optical depth (Goodman et al. 1995). This increase
would be expected if absorbing grains were aligned by magnetic field
within molecular clouds. 
On the other hand, far-infrared measurements (see Hildebrand 2000, henceforth
H00) 
showed strong polarization that is consistent with emission from
aligned grains. A quite general explanation to those facts was given
in Lazarian, Goodman \& Myers (1997, henceforth LGM), where it was argued that 
all the suspected alignment mechanisms are based on non-equilibrium processes
that require free energy to operate. Within the bulk of molecular clouds   
the conditions are close to equilibrium, e.g. the temperature difference
of dust and gas drops, the content of atomic hydrogen is substantially
reduced, and the starlight is substantially attenuated. As the result
the major mechanisms fail in the bulk part of molecular clouds apart
from regions close to the newly formed stars as well as the cloud
exteriors.

The alternative explanations look less appealing. For
instance, Wiebe \& Watson (2001) speculated that small scale turbulence
in molecular clouds can reduce considerably the polarization even if
grain alignment stays efficient. This, however, seems inconsistent with 
the results of 
far-infrared polarimetry (see H00) that
reveals quite regular pattern of magnetic 
field in molecular clouds. 

An extremely important study of alignment efficiency 
has been undertaken by Hildebrand and his
coworkers (Hildebrand et al. 1999, H00). They started by noticing that for a uniform
sample of aligned grains made of dielectric material consistent
with the rest of observational data $P(\lambda)$ should stay constant 
if $\lambda$ belongs to the far-infrared range. The data 
at 60 $\mu$m, 100 $\mu$m from Stockes on the Kuiper Airborne Observatory, 
350 $\mu$m from Hertz on Caltech Submillimeter Observatory, and
850 $\mu$m from SCUBA on the JCMT revealed a very different picture.
This was explained (see Hildebrand 2002) as the evidence for the
existence of populations of dust grains with different temperature
and different degree of alignment. The data is consistent with cold
(T=10~K) and hot (T=40~K) dust being aligned, while warm (T=20~K) grains
being randomly oriented (H00). 

If cold grains are identified
with the outer regions of molecular clouds, hot grains with regions
near the stars and warm with the grains in the bulk of molecular clouds
the picture gets similar to that in LGM. 
A quantitative theoretical study is absolutely necessary, nevertheless.
In LGM it was stated that 
within molecular clouds both
paramagnetic alignment aided by H$_2$ formation torques
and radiative torque alignment may be equally important. However,
it was later claimed in Lazarian \& Draine (1999) that the frequent
flipping of grains should make grain rotation essentially thermal (``
thermal trapping''). Recently calculated degrees of paramagnetic alignment
for thermally rotating grains
(Lazarian 1998, Roberge \& Lazarian 1999) that accounts 
for the complex grain dynamics are pretty low to explain the
observed degrees of polarization, however. This leaves the radiative
torque mechanism as the most probable candidate for alignment of dust
within molecular clouds. 
The quantitative testing
of the mechanism requires simulating radiative transfer within a
 molecular
cloud supplemented by a quantitative recipe for the alignment efficiency
dependence on the attenuated radiation spectrum. The latter element should
become available soon.

\subsection{Testing Alignment at the Diffuse/Dense Cloud Interface}

Grain alignment can be directly tested at the cloud interface. As we
mentioned earlier, Mathis (1986) explained the dependence of the polarization
degree versus wavelength , namely the Serkowski law (Serkowski 1973)
\begin{equation}
P(\lambda)/P_{max}=exp \left(-K ln^2(\lambda_{max}/\lambda) \right)~,
\end{equation}
(where $\lambda_{max}$ corresponds to the peak percentage polarization
$P_{max}$ and $K$ is a free parameter), 
assuming that it is only the grains larger than
the critical size that are aligned. Those grains were identified in 
Mathis (1986) as having superparamagnetic inclusions and therefore subjected
to more efficient paramagnetic dissipation. The ratio of the total to
selective extinction $R_v\equiv A_v/E_{B-V}$ reflects the mean size of 
grains present in the studied volume. It 
spans from $\sim 3.0$ in diffuse ISM to $\sim 5.5$ in dark clouds 
(see Whittet 1992 and references therein) as
the mean size of grain increases due to coagulation or/and mantle
growth. Earlier studies were consistent with the assumption that the
growth of $R_v$ is accompanied by the corresponding growth of 
$\lambda_{max}$ (see Whittet \& van Brenda 1978). The standard interpretation
for this fact was that as grains get bigger, the larger is the critical
size starting with which grains get aligned. This interpretation was
in good agreement with Mathis' (1986) hypothesis.

However, a recent study by Whittet et al. (2001) showed that grains
at the interface of the Taurus dark cloud do not exhibit the 
correlation of $R_v$ and $\lambda_{max}$. This surprising behavior
was interpreted in Whittet et al (2001) as the result of {\it
size-dependent variations in grain alignment} with small grains losing
their alignment first as deeper layers of the cloud are sampled. 
Whittet et al (2001) do not specify the alignment mechanism,
but their results pose big problems to the
superparamagnetic mechanism. Indeed, the data is suggestive that
$R_v$ and therefore the mean grain size may not grow  with extinction
while the 
critical size for grain alignment grows. The suprathermal
torques due to H$_2$ formation on grain surfaces (see Lazarian \& Draine
1997) do not look promising either, even if thermal trapping
(Lazarian \& Draine 1999a,b) is disregarded.
At the same time the observed behavior is  
exactly what is expected from radiative torques! Although the quantitative
comparison of the observations and the theoretical predictions is still
due to come, the results by Whittet et al. are very suggestive that the
radiative torques may be the dominant mechanism to align dust in ISM.  

\subsection{Testing Alignment for Small Grains}

Maximum entropy inversion technique in Kim \& Martin (1995) indicate that
 grains larger than a particular critical size are aligned. This is
consistent with our earlier discussion. However, an interesting feature of 
the inversion is that it is suggestive of smaller grains being partially
aligned. Initially this effect was attributed to the problems with the
assumed dielectric constants employed during the inversion, but a further
analysis that we undertook with Peter Martin indicated that the alignment
of small grains is real. Indeed, paramagnetic (DG) alignment must act on 
the small grains\footnote{To avoid a confusion we should specify that we
are talking about grains of $10^{-6}$~cm. For those grains the results
of DG relaxation coincide with those through resonance relation in
Lazarian \& Draine (2000). It is for grains of the size less than
$10^{-7}$~cm that the resonance relaxation is dominant.}. 
An important consequence of this is that the alignment
is proportional to the magnetic field strength. This opens an avenue for
a new technique of probing magnetic field using UV polarimetry\footnote{
UV polarimetry is sensitive to aligned small grains. As we discussed
earlier, the tiny PAH grains emit in microwave range, and their alignment
also depends on the magnetic field strength. Thus both microwave and
UV polarimetry may be useful in estimating the values of magnetic field.}. 

\section{Testing Alignment in New Environments}

{\it Indications of Alignment}\\
While interstellar grain alignment is an accepted process, alignment
of grains in other environments, e.g. comets, circumstellar disks, 
interplanetary medium, remains a controversial subject. Recent
advances in the alignment theory make us believe that 
grains are  well aligned there.
This is the point of view that was shared by a number of
earlier researchers. For instance, Greenberg (1970) claimed that
interplanetary dust should be mechanically aligned. Dolginov \& Mytrophanov
(1978) conjectured that comet dust and dust in circumstellar regions
was aligned. 
However, both the problems in understanding of grain alignment
and the inadequacy of  polarimetric data did not allow those
views to become prevalent (although see  Wolstencroft 1985, Briggs
\& Aitken 1986 where alignment was supported). 

As the situation with observations was gradually improving, the 
alignment of grains became difficult to disregard. 
It has been known for decades that 
various stars, both young and evolved, exhibit linear polarization
(see a list of polarization maps in Bastien \& Menard 1988). While
multiple scattering was usually quoted as the cause of the polarization,
recent observations indicate the existence of aligned dust around 
eta Carinae (Aitken et al. 1995),  evolved stars
(Kahane et al. 1997) and T Tauri stars (Tamura et al. 1999).
This is suggestive that for other stars grains should be aligned.
New observations (Chrisostomou et al 2000) support this. 
In fact, some of the arguments that
were used against aligned grains are favor them. For
instance, Bastien \& Menard (1988) point out that if polarization
measurements of young stellar object are interpreted in terms of grain
alignment with longer grain axes perpendicular to magnetic field,
the magnetic field of accretion disks around stars should preferentially
be in the disk plane. This is exactly what the
present day models of accretion disks suggest.

Similarly, 
``anomalies'' of polarization from comets\footnote{When light is scattered 
by the randomly oriented particles with sizes much
less than the wavelength, the scattered light is
 polarized perpendicular to the scattering plane, which is the
plane passing through the Sun, the comet and the observer. 
Linear polarization from comets has been long known to exhibit 
polarization that is not perpendicular to the scattering plane.} 
(see  Martel 1960, 
Beskrovnaja et al 1987, Ganesh et al 1998) as well as circular 
polarization\footnote{The subject was controversial for a while.
Observations of both left and right handed polarization in different
parts of coma with the average over whole coma close to zero probably
explains why earlier researchers were unsuccessful using large
apertures. Recent measurements by Rosenbush et al. (1999), Manset et al.
(2000) of circular polarization from Hale-Bopp Comet support the notion
that circular polarization is a rule rather than an exception.}
from comets ( Metz 1972, 
Metz \& Haefner 1987, Dollfus \& Suchail 1987, Morozhenko et al 1987) 
are indicative of grain alignment. 
However, conclusive measurements of grain alignment
have been done only recently for the Levi (1990 20) comet through direct
measurements of starlight polarization as the starlight was passing
through comet coma (Rosenbush et al 1994). The data proved the
existence of aligned grains, which corresponds to theoretical expectations
that we discuss below.

{\it Conditions for Alignment}\\
Expected measures of alignment and the necessary
conditions for the alignment to take place
are listed in Table~1. Both mechanical and radiative
alignment in comets and circumstellar regions 
can be transient, i.e. happen on the time scales less than
the Larmor precession time. My radiative torque
simulations with the code kindly supplied
to me by Bruce Draine show that when the magnetic field is not important
(i.e. $\tau_L\gg$ time of alignment),
the alignment tends to happen with grain angular momentum along the 
radiation flow. The rough estimate for the transient alignment time
is the time at which the angular momentum supplied either by radiative
torques or gaseous bombardment becomes comparable with the initial
grain angular momentum. The stationary alignment
requires a time larger than the gaseous damping time. For instance,
for rapid Larmor precession the radiative torque alignment on the
time scales much shorter than the gaseous damping time is marginal.
It can be seen from Table~1 that transient mechanical
alignment and that via radiative torques 
may act in opposite directions. If, as in the case of comets,
the radiation direction and the streaming direction coincide,
mechanical torques would tend to align grains parallel to the 
flow, while radiative torques will align grain perpendicular to
the flow. Accurate measurements of polarization direction may
determine the prevalence of one or the other mechanism. 
In the generally accepted picture of mechanical alignment, the increase of
angular momentum $J$ through collisions is proportional to $\sim t^{1/2}$, 
while radiative torques act to increase $J$ in proportion to time.
Thus even if mechanical alignment dominates initially, for grains
with $a\sim \lambda$ the radiative torques eventually dominate. The
mechanical alignment will still dominate for $a\ll \lambda$.

For mechanical alignment the existence of supersonic relative grain-gas
motions is essential. The joint action of radiative and mechanical torques
has not been studied yet, but we may suspect that the 
alignment may be 
caused by the difference in frictional damping parallel and perpendicular
to the flow (see Lazarian 1995). Models of stellar winds (see  
Netzer \& Elitzur 1994) and comet outflows predict supersonic relative
velocities and thus mechanical alignment. High resolution polarimetry
and modeling of the outflows are required.

High intensities of radiation flow make
radiative torques the most natural means of aligning grains
in circumstellar envelops. However, one should remember that
radiative torques are most efficient when grain sizes are comparable with
wavelengths. For particles much less than the wavelength their efficiency
drops as $(a/\lambda)^4$. This provides an interesting possibility that
large particles can be aligned via radiative torques, while small ones
may be aligned via other mechanisms. This effect may 
be revealed via spectropolarimetry of linear
and circularly polarized light. 
Note, that even in the interstellar medium, radiative torques are a 
major mechanism of rotation for sufficiently large, e.g. $a>10^{-5}$~cm,
grains. Within circumstellar regions, where UV flux is enhanced
smaller
grains can also be aligned radiatively. This could present a possible solution
for the recently discovered anomalies of polarization in the 2175 {\,{\rm \AA}}
 ~~extinction feature (see Anderson et al 1996) which have been interpreted
as evidence of graphite grain alignment (Wolff et al 1997). If
this alignment happens in the vicinity of particular 
stars with enhanced UV flux
and having graphite grains present in their circumstellar regions, this may
explain why no similar effect is observed along other lines of sight. 

We can see from Table~1 that if we assume a model for 
 Zodiacal dust particles that includes
 large silicate grains and small (less than the typical wavelength)
iron grains, both
species will be aligned with longer axes perpendicular to magnetic
field direction, although in the case of iron grains, the cause would
be paramagnetic relaxation, while in the case of silicate grains
the alignment would be due to radiative torques.

Potentially,
studies of Zodiacal Light might give a lot of information about magnetic
field structure and its variability in the Solar neighborhood, information
that could be compared with direct, in situ, measurements of the field. 
For instance,
the interplanetary magnetic field and fields in circumstellar regions
comet tails, are not stationary. In fact they undergo variations on a
whole range of time scales. If the variations are long compared to the
Larmor period $t_L$ they are adiabatic in the sense that the angle
between grain angular momentum and ${\bf B}$ is preserved. Therefore 
measuring  variations of the Zodiacal Light polarization
in a particular direction
can provide 
information on the magnetic
variability down to the scale $t_L$. 

\section{Discussion}

I anticipate a number of questions that can worry the reader. For instance:

$\bullet$
Does the review cover all the astrophysically important situations when
grain alignment is important? It has become clear recently that
grain alignment should happen in various astrophysical conditions.
Polarized radiation from neighboring galaxies (Jones 2000),
galactic nuclei (see Tadhunter et al 2001), 
AGNs, Seyfet galaxies (see Lumsden et al 2001), 
accretion discs (see Aitken et al 2002) can be partially due to
aligned particles. Revealing this contribution would allow to study
magnetic fields in those and other interesting objects. 

$\bullet$
To what degree do aligned grains reveal magnetic field geometry/topology
during star formation? It is generally accepted that star formation
starts with the accumulation of interstellar gas that is
caused by turbulence and gravity. Aligned grains allow to trace magnetic
fields during this preliminary stage. At some point of evolution the conditions
within molecular clouds approach equilibrium with the alignment being shut
down (see LGM). Finding out exactly when this happens is extremely important
and this requires the quantitative description of grain alignment processes.
Consider, for instance, radiative torques. Realistic 
clumpy, fractal-type structure
of molecular clouds allows photons to penetrate much deeper into clouds
compared with the idealized uniform structure frequently assumed in
theoretical modeling. Therefore we expect grains within skin layers of
the clumps to be aligned and to reveal magnetic field up to a substantial
optical depth. Simulations in (Padoan et al. 2001) support this argument.
As  protostars are formed in molecular clouds their light induces
grain alignment in their neighborhood. The size of this neighborhood also
depends on the cloud inhomogeneity in the protostar vicinity
as well as on the radiative torque efficiency as a function of wavelength.
The fact that grains in molecular clouds are larger than their 
counterparts in diffuse media
 allows for a more efficient alignment by starlight reddened
by dust extinction; this increases the neighborhood volume. Therefore
we expect to
be able to trace magnetic evolution via polarimetry
through important stages of star
formation. Additional information can be available through microwave
emission of the aligned PAH-type tiny grains, which rotate non-thermally
due to their collisions with ions (see Draine \& Lazarian 1998b). The
abundance of such grains in molecular clouds is poorly known, however.

$\bullet$
What is the advantage of far-infrared polarimetry for studies of magnetic
field in molecular clouds compared to optical and
near infrared ones? The trivial answer is that far infrared
polarimetry reveals aligned grains near newly born stars unaccessible
by optical and near infrared photons. A more subtle but essential effect is 
that photons, as we discussed earlier,
 can align grains within skin layers of clumps rather far into molecular
 clouds. Those aligned grains are only accessible by 
far infrared polarimetry. This, for instance, makes SOFIA airborn observatory
so desirable for studies of magnetic fields in molecular clouds.
Additional advantage of far infrared 
spectropolarimetry  stems from the fact that it allows us
to separate contributions from different parts of the cloud (see Hildebrand
2000). This
enables tomography of magnetic field structure. 

$\bullet$
What is the future of optical and near infrared polarimetry?
It would be wrong to think that with the advent of far infrared
polarimetry there is a bleak future for extinction polarimetry at
shorter wavelengths. In fact, its potential for studies
of magnetic fields in the Galaxy is enormous (see Fosalba et al.
2002, Cho \& Lazarian 2002a). The possibility of using
stars at different distances from the observer allows to get an insight
into the 3D distribution of magnetic fields. In general, however,
it is extremely advantageous to combine optical/near infrared and
far infrared polarimetry. For instance, it may be pretty challenging
to trace the connection of Giant Molecular Clouds (GMCs) with the
ambient interstellar medium using just far-infrared measurement.
However, if extinction polarimetry of the nearby stars is included,
the task gets feasible. Similarly testing modern concepts of MHD
turbulence (Goldreich \& Shridhar 1995, Lithwick \& Goldreich 2001,
Cho \& Lazarian 2002b) 
and turbulent cloud support (see reviews by 
McKee 1999, and Cho, Lazarian \& Vishniac
2002) would require data from both  diffuse and dense media. 

$\bullet$
What is the advantage of doing polarimetry for different wavelengths?
The list of advantages is pretty long. It is clear that
aligned grains can be successfully used as pick up
devices for various physical and chemical processes, provided
that we understand the causes of alignment. Differences in alignment
of grains of different chemical composition (see Smith et al. 2000)
provides a unique source of the valuable information. Comets present
another case in support of simultaneous multifrequency
studies. There the
properties of dust evolve in a poorly understood
fashion and this makes an unambiguous interpretation
of optical polarimetry rather difficult. Degrees and directions
of dust alignment that can be obtained that can be obtained via 
far infrared polarimetry can be used to get a self-consistent
picture of the dust evolution and grain alignment.   

$\bullet$
Do we need grain alignment theory to deal with polarized CMB foregrounds?
Polarized emission spectrum arising from aligned dust may be very complex
if grains having different temperatures exhibit different degrees of 
alignment. In this situation the use of the naive power-law templates 
may result in huge errors unless we understand grain alignment properly.
Needless to say that grain alignment theory is necessary to predict the
spectrum of polarized emission from PAHs in the range of 10-100~GHz.

$\bullet$
What is the future of grain alignment theory? Although the recent
progress in understanding grain alignment is really 
encouraging, it would be
a mistake to think that grain alignment theory does not require 
intensive work any more. For instance, radiative torques alignment in the
presence of starlight anisotropy 
should be treated as an experimental fact obtained
via simulations rather than a theoretically understood effect. Moreover,
crossover dynamics must be added to the existing code
to get the simulations more
realistic and frequency dependence of radiative torques should
be quantified. More special cases of alignment should be studied.
The simultaneous action of various processes, e.g. grain streaming
together with the action of radiative torques must be investigated. Some
additional processes, e.g. mechanical alignment of helical grains (see
table~1. in LGM) must be quantified. Alignment of tiny PAH
grains, in particular, is an essentially unexplored field that requires
more studies of relaxation processes in minute quantum mechanical samples
as well as plasma and magnetic turbulence interactions with grains.
More observational testings are necessary as well. For instance, comets
allow to trace grain alignment in time. More systematic studies that
include not only linear polarimetry, but circular polarimetry as well,
should be made. All in all, grain alignment has become a predictive theory,
but there is more work, both observational and theoretical to be done.

\section{Summary}

The principal points discussed above are as follows:
\begin{itemize}
\item Grain alignment results in linear and circular polarimetry.
The degree of polarization depends on the degree of grain alignment
and the latter is the subject of grain alignment theory.
\item Grain alignment theory has at last reached its mature state
when predictions are possible. In most cases grain alignment happens
in respect to magnetic field, i.e. reveal magnetic field direction,
even if the alignment mechanism is not
magnetic. However, depending on the mechanism the alignment may
happen with grain longer axes parallel or perpendicular to magnetic
field.  
\item Radiative torques, after being ignored for many years, have become
the most promising mechanism which expectations agree with interstellar
observations. However, astrophysical circumstances exhibit such a variety
of conditions that other mechanisms have their own niches.
\item Advances in grain alignment theory make it possible to use
grains as sensitive pick up devices. For instance, we discussed
ways to use polarization as a direct measure of magnetic field intensity. 
\item It is clear that the importance of grain alignment is not limited
to interstellar medium and molecular clouds. Polarimetry can be used
to study magnetic fields in accretion disks,
AGN, circumstellar regions, comets etc.
\end{itemize}

Support by NSF grant AST-0125544 is acknowledged. It is a great pleasure to
thank Bruce Draine, Roger Hildebrand and John Mathis for many discussions.
Communications with many of my colleagues, especially with Alyssa Goodman
were very helpful. Comments by Dave Aitken improved the manuscript.

\begin{figure}[h]
\includegraphics[width=2in]{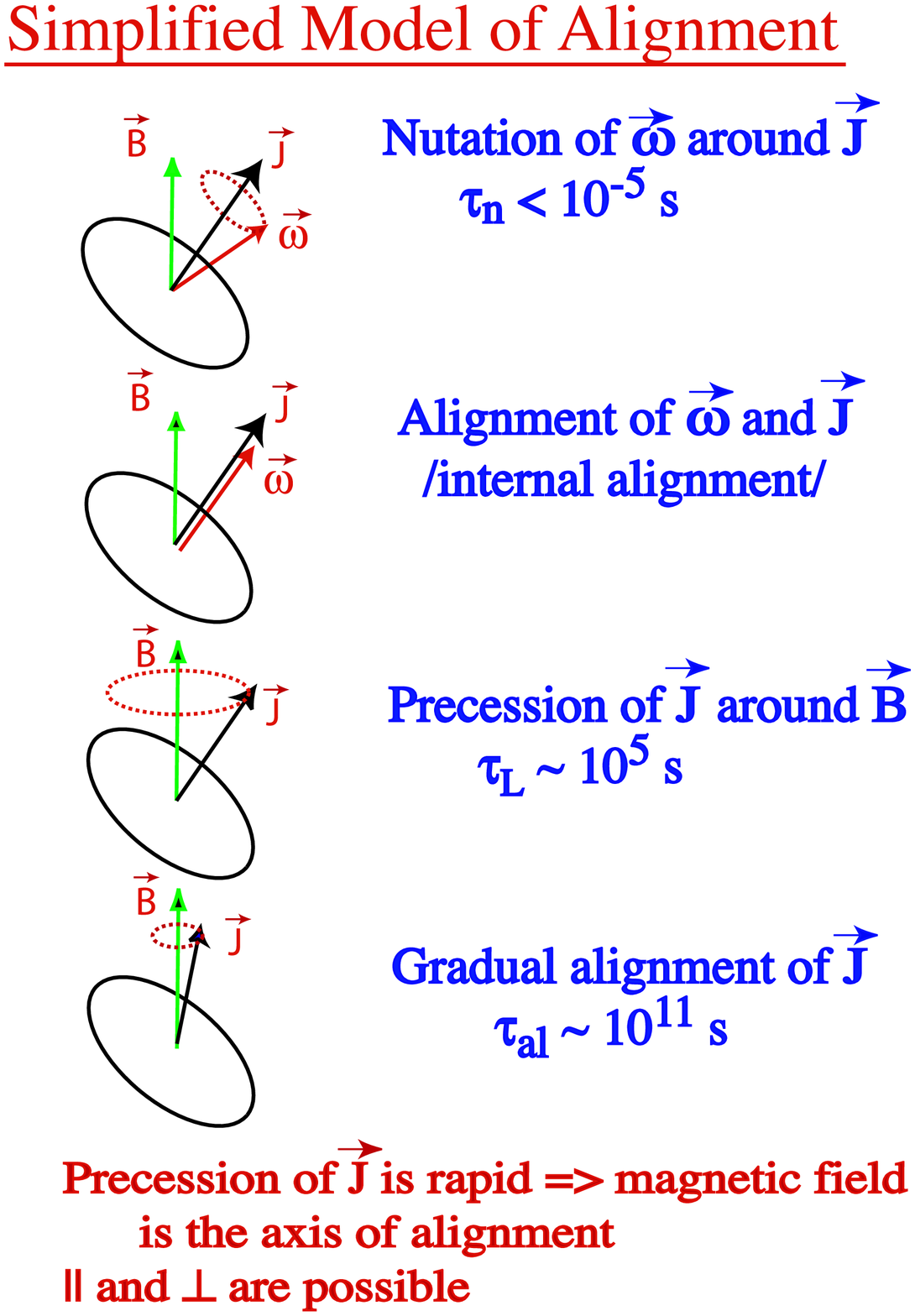}
\hfill
\includegraphics[width=3.in]{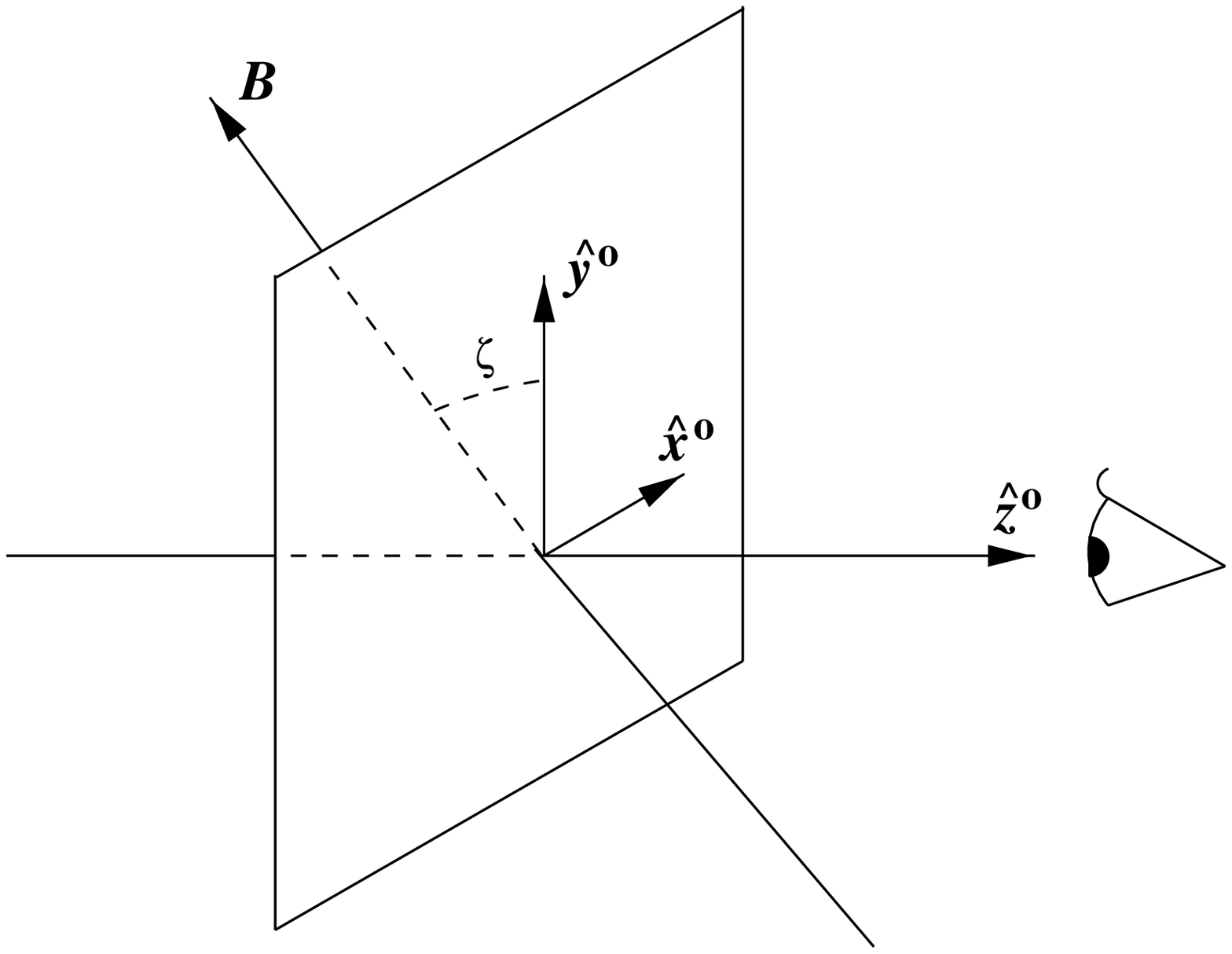}
\caption{a)Left panel. Alignment of grains implies several alignment
processes acting simultaneously spaning over many scales. 
Internal alignment was introduced by Purcell (1979) and was assumed to be
a slow process. Lazarian \& Draine (1999a) showed that the internal alignment
is $10^6$ times faster if nuclear spins are accounted for.  The time
scale of ${\bf J}$ and ${\bf B}$ alignment is given for diffuse interstellar
medium. It is faster in circumstellar regions and for comet dust.
 b) Right panel.
Geometry of observations (after Roberge \& Lazarian 1999).} 
\label{fig:2Dspek}
\end{figure}

\begin{table}[h]
\begin{displaymath}
\begin{array}{lrrrr} \hline\hline\\
& \multicolumn{1}{c}{\it Features} & \multicolumn{1}{c}{\bf Comets} &
\multicolumn{1}{c}{\bf Circumstellar} {\bf Regions}\\[1mm]
\hline\\
{\rm \bf Radiative}& {\it options}:& {\rm transient~and~stationary}
& {\rm stationary~mostly}\\[1mm]
{\rm \bf Torques}&{\it conditions}: & {\rm 
effective~when}~a\sim\lambda & 
{\rm the~same} \\[1mm]
 &{\it direction}: & \bot~{\rm to}~{\rm photon~flux}~{\rm for~transient}& \bot~{\rm to~magnetic~field}~{\bf B}\\[1mm]
  & & \bot~{\rm to}~\bf B~{\rm for~stationary~alignment} & \\[1mm]
  &measure:& R\sim 1 & R\sim 1\\[1mm]
\hline \\[1mm]
{\rm \bf Paramagnetic}&{\it conditions:}& {\rm 
pure~iron~grains} & {\rm grains~with~
inclusions}\\[1mm]
{\rm \bf Alignment}& {\it direction:}& \bot~{\rm to}~{\bf B}& 
{\rm the~same} \\[1mm]
& measure:& {\rm up to}~R\sim 1 &  
{\rm the~same} \\[1mm]
\hline\\
{\rm \bf Mechanical}&{\it options:}&{\rm transient~and~stationary}& 
{\rm mostly~stationary}\\[1mm] 
{\rm \bf Alignment}& {\it conditions:}& {\rm supersonic~drift}& 
{\rm the~same}  \\[1mm]
&{\it direction}&\| {\rm 
to~gas~flow~for~transient}&
\bot~{\rm or}~\|~{\rm to}~\bf B\\[1mm]
&& \bot~{\rm or}~\|~{\rm to}~\bf B~{\rm for~stationary}& \\[1mm]
&measure:& {\rm from}~-0.5~{\rm to}~0.4& {\rm the~same}
 \\[1mm]\hline
\end{array}
\end{displaymath}
\caption{Conditions for successful alignment and the expected measures
of the Rayleigh reduction factor $R$. For
paramagnetic alignment only the {\it stationary} option is available.
For grains with $a\ll \lambda$ the radiative torques are negligible.} 
\label{tab:2Dspk_asymp}
\end{table}

\end{document}